\documentclass[12pt]{article}
\usepackage{epsfig}
\catcode`\@=11
   
\long\def\@makefntext#1{\parindent 0cm\noindent
\hbox to 1em{\hss$^{\@thefnmark}$}#1}

\begin{document}

\begin{titlepage}
\vspace{.5in}
\begin{flushright}
UCD-02-14\\
hep-th/0209249\\
September 2002\\
\end{flushright}
\vspace{.5in}
\begin{center}
{\Large\bf
 Black holes may not\\[.5ex] constrain varying constants}\\
\vspace{.4in}
{S.~C{\sc arlip}\footnote{\it email: carlip@dirac.ucdavis.edu}
and {S.~V{\sc aidya}\footnote{\it email: vaidya@dirac.ucdavis.edu}}\\
       {\small\it Department of Physics}\\
       {\small\it University of California}\\
       {\small\it Davis, CA 95616}\\{\small\it USA}}
\end{center}

\vspace{.5in}
\begin{center}
{\large\bf Abstract}
\end{center}
\begin{center}
\begin{minipage}{4.7in}
{\small New and rather controversial observations hint that the fine
structure constant $\alpha$ may have been smaller in the early
universe, suggesting that some of the fundamental ``constants'' of
physics may be dynamical.  In a recent paper \cite{Davies}, Davies,
Davis, and Lineweaver have argued that black hole thermodynamics 
favors theories in which the speed of light $c$ decreases with time, 
and disfavors those in which the fundamental electric charge $e$
increases.  We show that when one considers the full thermal
environment of a black hole, no such conclusion can be drawn:
thermodynamics is consistent with an increase in $\alpha$ whether it
comes from a decrease in $c$, an increase in $e$, or a combination of
the two.}
\end{minipage}
\end{center}
\end{titlepage}
\addtocounter{footnote}{-2}

Recent observations of spectral lines of distant quasars have
suggested that the fine structure constant $\alpha = e^2/\hbar c$ may
have been slightly smaller in the very early universe \cite{Webb}.
Although these claims are still tentative and rather controversial,
they have helped rekindle interest in Dirac's old idea \cite{Dirac}
that the fundamental ``constants'' of physics may vary in time.  In a
recent Brief Communication to {\it Nature\/}, Davies, Davis, and
Lineweaver have argued that black hole thermodynamics favors theories
in which the speed of light $c$ decreases in time, and disfavors those
in which the fundamental electric charge $e$ increases
\cite{Davies}. We show here that when one considers the full thermal
environment of a black hole, no such conclusion can be drawn.

The fine structure constant depends on the fundamental charge $e$,
Planck's constant $\hbar$, and the speed of light $c$, and it is
natural to ask which of these varies.\footnote{Variation of
dimensionful constants is inherently ambiguous \cite{Duff}; here, we
take ``varying $e$'' to mean ``suitable variation of all dimensionless
quantities that depend on $e$.''}  Davies et al.\ offer an ingenious
argument.  A black hole with mass $M$ and charge $Q=ne$ has an entropy
\cite{Hawking}
\begin{equation}
S/k = \frac{\pi G}{\hbar c}\left[ M + \sqrt{M^2 - n^2e^2/G}\right]^2
\label{a1}
\end{equation}
Evidently an increase in $e$ will decrease this entropy, apparently
violating the generalized second law of thermodynamics, while a
decrease in $\hbar$ or $c$ will increase the entropy.

As Davies et al.\ point out, though, such entropic considerations
should take into account not just the black hole, but its surroundings
as well. An isolated black hole is never in thermal equilibrium: it
will always decay by Hawking radiation and, if it is charged, by
Schwinger pair production \cite{Gibbons} as well.  These processes
decrease $S$, but they do not violate the second law, since the
decrease is compensated by an increase in the entropy of the
environment.

To study the full thermodynamics of ``varying constants,'' one may try
to examine the detailed dynamics of heat flow and entropy between a
black hole and its environment.  A simpler alternative is to consider
a proper thermodynamic ensemble, that is, a black hole in a heat bath.
To obtain an equilibrium, one must consider a black hole in a ``box''
of radius $r_B$, with fixed temperature $T$ and charge $Q$ (canonical
ensemble) or fixed temperature $T$ and electrostatic potential $\phi$
(grand canonical ensemble) at the boundary.  For charged black holes,
such ensembles were first studied by Braden et al.\ \cite{Braden}.  In
the canonical ensemble, the total entropy takes the form $S = \pi
r_B{}^2x^2$, where $x$ is determined by the seventh-order equation
\begin{equation}
x^5 (x-q^2)(x-1) + b^2 (x^2 - q^2)^2 = 0
\label{a2}
\end{equation}
with $q = \sqrt{G}Q/r_Bc^2$ and $b = \hbar c/4\pi r_B kT$.

\begin{figure}
\centerline{\epsfig{figure=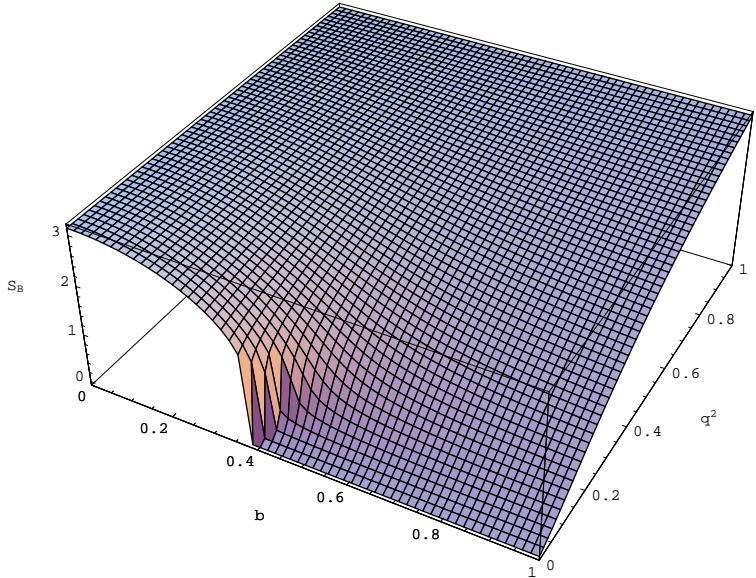,height=3.1in}}
\caption{Entropy as a function of $q^2$ and $b$.}
\label{fig:entropy}
\end{figure}

While no analytical solution of (\ref{a2}) is possible for generic
charge and temperature, a numerical solution is straightforward.
Figure 1 shows a graph of $S_B \equiv S/r_B^2$ against $q^2$ and $b$,
restricted to the highest-entropy configuration.  It is apparent---and 
may be confirmed with more detailed numerical study---that the 
total entropy {\it increases\/} with increasing $q^2$, and thus with 
increasing $\alpha$.  For the grand canonical ensemble, exact analytic 
results can be found, and the conclusion is the same.  Black hole
thermodynamics thus militates against models in which the 
fundamental charge $e$ decreases, but places no restrictions
on models in which $e$ increases.

How can one reconcile this result with the argument of Davies et al.?
Note first that the Hawking temperature,
\begin{equation}
kT_H = \frac{\hbar c^3}{8\pi G}
  \frac{\sqrt{M^2 - n^2e^2/G}}{2M^2 - n^2e^2/G + 2M\sqrt{M^2 - n^2e^2/G}}
\label{a3}
\end{equation}
decreases with increasing $e$.  As $e$ increases, a black hole will thus cool 
below the ambient temperature of the heat bath, and will absorb heat, 
increasing its mass.  By the first law of thermodynamics, the net change in
entropy is
\begin{equation}
dS = \frac{1}{T} \left( dE - \phi dQ\right) 
\label{a4}
\end{equation}
and it may be checked explicitly that the increase in quasilocal
energy $E$ more than compensates for the direct effect of changing
$Q$.

Of course, thermodynamic arguments of this sort only describe
relationships among equilibria, and not the details of the transitions
between equilibria.  In some ways, this is an advantage: our results
are rather insensitive to the details of a theory describing ``varying
constants,'' requiring only that conventional black hole thermodynamics
hold at equilibria at which $\alpha$ is constant.  On the other hand, this
generality allows some loopholes: one could, for instance, imagine a 
scenario in which an abrupt change in $\alpha$ led to an initial decrease 
in entropy, which then grew only slowly as heat was redistributed.  

To analyze such possibilities, though, there is no substitute for a
detailed dynamical model.  In particular, any theory with a variable
fine structure constant necessarily contains a new scalar field,
$\alpha$ itself.  The entropy of that field might be negligible at
equilibrium, but it surely cannot be ignored during dynamical
processes in which $\alpha$ is changing.  Black hole mass quantization
may constrain models with ``varying constants'' \cite{Carlip}, but in
the absence of a detailed dynamical description, it seems that black
hole thermodynamics cannot.

\vspace{1.5ex}
\begin{flushleft}
\large\bf Acknowledgments
\end{flushleft}

\noindent This work was supported in part by U.S.\ Department of Energy 
grant DE-FG03-91ER40674.

\end{document}